\documentclass[aip, jcp, reprint, floatfix]{revtex4-1}   

\usepackage{graphicx}
\usepackage{dcolumn}
\usepackage{bm}

\usepackage[utf8]{inputenc}
\usepackage{mathptmx}
\usepackage{etoolbox}
\usepackage{subcaption}
\usepackage[version=4]{mhchem}
\usepackage{siunitx}
\usepackage{mhchem}
\usepackage{placeins}
\DeclareSIUnit\angstrom{\text {Å}}

\makeatletter
\def\@email#1#2{%
 \endgroup
 \patchcmd{\titleblock@produce}
  {\frontmatter@RRAPformat}
  {\frontmatter@RRAPformat{\produce@RRAP{*#1\href{mailto:#2}{#2}}}\frontmatter@RRAPformat}
  {}{}
}%
\makeatother

\begin{document}
\preprint{AIP/123-QED}

\title{Paying attention to long-range electron correlation: a size-independent deep-learning approach to predicting molecules' electronic energies from one- and two-electron integrals}


\author{Valerii Chuiko}
\affiliation{ 
Chemistry and Chemical Biology, McMaster University, Hamilton, Ontario, L8S 4L8, Canada
}

\author{Giovanni B. Da Rosa}
\affiliation{Télécom Paris, Institut Polytechnique de Paris, Palaiseau, France}
\affiliation{Instituto de Física, Universidade Federal do Rio Grande do Sul (UFRGS), Porto Alegre, RS, Brazil}

\author{Paul W. Ayers}%
 \email{ayers@mcmaster.ca}
\affiliation{ 
Chemistry and Chemical Biology, McMaster University, Hamilton, Ontario, L8S 4L8, Canada
}


\date{\today}

\begin{abstract}
We propose a descriptor for molecular electronic structure that is based solely on the one- and two-electron integrals but is translationally, rotationally, and unitarily invariant. Then, directly exploiting size consistency, we train and fine tune a neural network to predict the energies of strongly-correlated systems, specifically hydrogen clusters. We use an attention mechanism to formulate a size-independent approach that uses and preserves size-consistency. Therefore, training on few-electron systems can guide predictions for systems with more electrons. Our results are more accurate than alternative geometry-based machine-learning models.

\end{abstract}

\maketitle

Modeling the electronic structure of substances is essential in many scientific disciplines, including materials science\cite{Rastogi1993, Lin2014, Armstrong2006}, pharmaceuticals\cite{Arodola2017, Fogueri2013, Gordon2007}, and catalysis\cite{Chen2017, Chen2010}. At its most fundamental level, this requires determining the electronic wavefunction, density matrix, or a descriptor that suffices to determine these quantities. Using the wavefunction/density matrix, researchers can then predict all observable molecular properties, including static properties (e.g., energy levels) and (thermo)dynamic properties (e.g., reaction kinetics).  

Determining the electronic wavefunction requires solving the Schr\"{o}dinger equation. The exact solution of Schr\"{o}dinger equation (within the function space defined by a chosen one-electron basis set) can be calculated using the Full Configuration Interaction (FCI) method. This method represents the wave function as a linear combination of all possible configurations of the system's electrons. However, FCI suffers from the curse of dimensionality\cite{Bellman1957}, which makes it computationally intractable for the vast majority of chemically-relevant systems. Consequently, traditional wave function methods rely on approximations and assumptions that limit their accuracy and applicability. 

\begin{raggedright}
In recent years, machine learning (ML) and, specifically, neural networks (NN), have emerged as powerful tools for predicting the energy of chemical systems, offering an efficient alternative to traditional electronic structure approaches\cite{Fabrizio_Meyer_Fabregat_Corminboeuf_2019, Muroga2023, Jinich, Schtt2017,Schtt2019, No2020, Clemence2022}. These techniques can capture complex relationships within molecular data and generalize well to unseen molecules, significantly accelerating the computation of molecular energies and other properties. For example, neural networks can be trained to learn potential energy surfaces\cite{Schtt2017, SchNet, Schütt2020, Clemence2022, Skala}, IR and Raman Spectra\cite{Muroga2023}, force fields in molecular dynamics simulations\cite{Brockerde2017, Zhang2018, Zeng2023}, and even energies\cite{Mazziotti2022, Skala}, wavefunctions\cite{Schtt2019} and reduced density matrices\cite{GBooth2018, Pavanello2023, Mazziotti2025}. Advanced architectures like deep neural networks (DNNs), convolutional neural networks (CNNs), and graph neural networks (GNNs) have been applied to model molecular properties, leveraging vast datasets to improve their predictive capabilities.
\end{raggedright}

However, ML and deep learning (DL) methods have limitations. They require large amounts of high-quality data for training, which can be difficult to obtain, especially for less-studied systems or those involving rare or complex chemical species\cite{Dral2020}. Additionally, these models can be prone to overfitting: they often perform well on training data but poorly on new, unseen, data\cite{Dral2020, Huang2021}. Despite these challenges, ML and DL are increasingly being integrated into the workflows of computational chemists and materials scientists, facilitating the discovery and design of new materials and drugs.

The aim of this work is to address some of the most significant challenges associated with current deep learning methods for predicting electronic energies. Specifically, we focus on the need for (a) high-quality datasets, (b) rotational and translational invariance of the input features, and (c) transferability to new systems. These issues are critical for the development of reliable and generalizable DL models that can accurately predict the properties of a wide range of systems. Our methodology combines costly data curation (from FCI calculations), innovative featurization, and advanced recycling of available data.
In the first part of this paper we introduce a descriptor of the electronic Hamiltonian that is rotationally, translationally, and unitarily invariant. Using this descriptor, every learned FCI sample can be used to predict the energy of an infinite number of chemically-equivalent systems. This descriptor also ensures that the mapping from the electronic Hamiltonian to the ground-state electronic energy has the correct mathematical properties. In the second part of this paper we use our descriptor to predict molecules' electronic energy using deep learning methods.

As stated above before, our first task is to develop a descriptor that is invariant to changes in the representation of the system. Our descriptor will be based on the one- and two-electron integrals that define the Hamiltonian in second quantized form:
\begin{equation}
\hat{H}=\sum_{p q} h_{p q} a_p^{\dagger} a_q+\frac{1}{2} \sum_{p q r s} V_{p q r s} a_p^{\dagger} a_q^{\dagger} a_s a_r
\end{equation}
where
\begin{subequations}
\begin{align}
    h_{pq} &= \int{\phi_p^{*}(\mathbf{r}) \left(-\frac{\nabla^2}{2} - \sum_{n=1}^{N_\text{atoms}}{\frac{Z_n}{|\mathbf{r} - \mathbf{R}_n|}}\right) \phi_q(\mathbf{r})}\,d\mathbf{r} \\
    V_{pqrs} &= \iint{\phi_p^{*}(\mathbf{r}_1) \phi_q^{*}(\mathbf{r}_2) \frac{1}{\left|\mathbf{r}_1 - \mathbf{r}_2\right|} \phi_r(\mathbf{r}_1) \phi_s(\mathbf{r}_2)}\,d\mathbf{r}_1 d\mathbf{r}_2
\end{align}
\end{subequations}
and $a_p^{\dagger}$ and $ a_p$ denote the fermion creation and annihilation operators for an electron in the p\textsuperscript{th} spin-orbital. It is convenient express the one-electron operators as two-electron operators, so that the $N$-electron Hamiltonian can be described with only one type of term:
\begin{equation}
\label{reduced_ham}
\begin{split}
\hat{H}_N &= \sum_{p q r s} \left[\frac{1}{2(N-1)} \left( h_{pq} \delta_{rs} + \delta_{pq} h_{rs} \right) + \frac{1}{2} V_{p q r s} \right] a_p^{\dagger} a_q^{\dagger} a_s a_r \\
&= \sum_{p q r s} k_{pqrs} a_p^{\dagger} a_q^{\dagger} a_s a_r 
\end{split}
\end{equation}

To enforce unitary invariance, we recall that the only unitarily invariant properties of a matrix are its eigenvalues. This motivates us to rewrite the 4-dimensional tensor $k_{pqrs}$ as a matrix. To do this, we project the Hamiltonian onto a geminal basis, where geminal creation and annihilation operators are defined as:

\begin{subequations}
\begin{align}
    g_A^{\dagger} & =\frac{1}{\sqrt{2}} a_p^{\dagger} a_q^{\dagger} \\
    g_A & =\frac{1}{\sqrt{2}} a_q a_p
\end{align}
\end{subequations}

Unitary rotation of the geminal encapsulates unitary rotation of the orbitals.\cite{sørensen2023transformationgeminalbasisstationary} Therefore, the eigenvalues of the Hamiltonian matrix elements in the geminal basis, $k_{AB}$, are the unitarily invariants of the Hamiltonian and therefore contain enough information to determine its eigenvalues. The eigenvalues of $k_{AB}$ are rotationally, translationally, and unitarily invariant, so they are a suitable input for a traditional feed-forward neural network, as discussed below.

We additionally benchmark our results against geometry-based descriptors. A closely related descriptor reported in the literature is built from permutationally invariant polynomials (PIPs). To generate the PIPs descriptors we used the MOLPIPx\cite{MOLPIPx} software package. 

In MOLPIPx the corresponding feature vector is defined as
$$
\boldsymbol{\Phi}_{\mathrm{PIP}}(\mathbf{x})=\left(\mathrm{f}_{\text {poly }} \circ \mathrm{f}_{\text {mono }} \circ \gamma \circ d\right)(\mathbf{x}),
$$

where each function in this sequence applies a nonlinear transformation:

$$
\begin{aligned}
\mathbf{r} & =d(\mathbf{x}), & & \text { interatomic distances } \\
\bar{\gamma} & =\gamma(\mathbf{r})=e^{-\lambda \mathbf{r}}, & & \text { Morse-type variables } \\
\mathbf{z}_{\text {mono }} & =\mathrm{f}_{\text {mono }}(\bar{\gamma}), & & \text { symmetrized monomials } \\
\mathbf{z}_{\text {poly }} & =\mathrm{f}_{\text {poly }}\left(\mathbf{z}_{\text {mono }}\right), & & \text { symmetrized polynomials }.
\end{aligned}
$$
Here, $\lambda$ is a length-scale hyperparameter, while the mappings $f_{\text {mono }}$ and $f_{\text {poly }}$ are generated using the monomial symmetrization \cite{PIPs}. We used fourth-order polynomials to 
generate features. 
 
We also tested our method against SchNet\cite{SchNet}, which we retrained on the same datasets to ensure a fair comparison. In the last part of the paper we tested our approach against the recently presented Skala\cite{Skala} model. Because Skala already incorporated training data from hydrogen clusters, We did not retrain Skala for our specific dataset.

\begin{raggedright}
 We chose hydrogen clusters as test systems because they are extremely challenging for many quantum-mechanical models because of the presence of strong electron correlation, yet computationally tractable for benchmark FCI calculations. Specifically, we examine different geometric families of neutral 2-, 4-, 6-, 8-, and 10-atom hydrogen clusters. For each system we performed FCI calculations with the STO-6G basis set using PySCF\cite{sun2018pyscf}.
\end{raggedright}
 
\textbf{\ce{H2}; 156 geometries}:
 We sampled the potential energy curve of \ce{H2} molecule for bond lengths between 0.2 \si{\angstrom} and 8 \si{\angstrom}. 
 
\textbf{\ce{H4}; 865 geometries}:
 We selected three families of \ce{H4} structures: Paldus's \ce{H4} system \cite{Li1995}, stretching linear \ce{H4}, and tetrahedral inversion. As before, the linear \ce{H4} stretch included interatomic separations from 0.2 \si{\angstrom} to 8 \si{\angstrom}. The tetrahedral family represented an inversion of a tetrahedron, with interatomic distances ranging from 1 \si{\angstrom} to 5 \si{\angstrom}. Paldus's \ce{H4} family\cite{Li1995} spanned interatomic distances from 0.5 \si{\angstrom} to 5 \si{\angstrom}. 

\textbf{\ce{H6}; 1386 geometries}:
 We selected three families of \ce{H6} structures: planar symmetry breaking, a triangular antiprism, and octahedral distortions. To study symmetry-breaking, we overlapped two equilateral triangles in the same plane to form a regular hexagon. We then rotated one of the triangles from 0{\textdegree} to 30{\textdegree}; we repeated this for equilateral triangles with side-lengths from 2  \si{\angstrom} to 6  \si{\angstrom}. The triangular antiprism family captures the evolution from a triangular prism to a triangular antiprism by rotating one of the bases. In the initial triangular prism, the interplanar separation is the same as the side-length of the triangle, which varies from 1 \si{\angstrom} to 4 \si{\angstrom}. In the final triangular antiprism, the side length of the triangles is the same as the distance of each atom to the closest atoms in the other triangle. We made a linear interpolation between these two structures. The family of octahedral distortions starts from a regular octahedron, then the axial atoms are pushed towards the center. The octahedral side length ranges from 2 \si{\angstrom} to 8 \si{\angstrom}. 

\textbf{\ce{H8}; 362 geometries}:
We selected five families of \ce{H8} structures: Paldus's \ce{H8} system\cite{Li1995}, hydrogen chain stretching, the M{\"o}bius-Kantor polygon, the square antiprism, and planar symmetry breaking. Paldus's \ce{H8} structures are generated as described in the original paper\cite{Li1995}.
The hydrogen chain stretch covers interatomic distances from 0.2 \si{\angstrom} to 6 \si{\angstrom}.
The M{\"o}bius-Kantor polygon is a square with each edge capped by a hydrogen atom. The inner atoms move along perpendiculars to each edge, with interatomic distances ranging from 1 \AA to 4 \AA.
The square antiprism, similar to the \ce{H6} triangular antiprism, was calculated only for the evolution from antiprism to prism at an interatomic distance of 2 \si{\angstrom}. Analogous to \ce{H6}, planar symmetry breaking was studied by stacking two squares with side-length 2 \si{\angstrom} on top of each other in the plane to form a regular octagon, and then rotating one of the squares.

\textbf{\ce{H10}; 150 geometries}:
For \ce{H10} we exclusively considered the dissociation of a hydrogen chain from 0.5 \si{\angstrom} to 8 \si{\angstrom}.

To test our approach, we trained three different neural networks to predict electronic energies of 4, 6, and 10-atom hydrogen clusters. We used the Adam optimizer for training with a learning rate of $10^{-3}$.

For all systems we used only fully connected dense layers. For predicting the energy of 4 and 6 hydrogen atoms we used one layer with 100 neurons with the ReLu activation function, followed by another layer with 50 neurons with the sigmoid activation function. The final, output, layer used a single neuron with a linear activation function. To compare our approach to PIPs we generated descriptors using the MOLPIPx\cite{MOLPIPx} library and trained exactly the same NN. In case of SchNet, we used the original architecture. 

For the training data, we randomly selected 80\% of the systems, while the remaining 20\% were used to evaluate the model's performance. The learning curves for the training procedure are shown in the supplementary information.



In Table 1 we report the prediction errors of our model on the test set and compare it to the results of several popular quantum chemistry methods: Hartree Fock (HF), CCSD(T)\cite{RAGHAVACHARI1989479}, B3LYP\cite{Becke1993, Lee1988}, and MP2\cite{MP2}. The best quantum chemistry methods have \textit{very large} errors $\sim .3$ a.u., while our neural networks has errors $\sim .002$ a.u., approaching chemical accuracy. While SchNet and MOLPIPx produce the similarly small errors. However, as we shall see later, they are not generalizable.

\begin{table}[htbp]
\begin{tabular}{|l|c|c|c|c|c|c|c|}
\hline
 & \multicolumn{7}{c|}{MAE (a.u.)} \\
\hline
System & HF & CCSD & MP2 & B3LYP & SchNet & MOLPIPx & NN \\
\hline
H4 & 0.5890 & 0.3154 & 0.4128 & 0.6278 & 0.0041 & 0.0039 & 0.0022 \\
\hline
H6 & 0.5817 & 0.3716 & 0.2793 & 0.3983 & 0.0046 & 0.0035 & 0.0024 \\
\hline
\end{tabular}
\caption{Energy errors for the test set for different methods}
\end{table}

\begin{figure}[htbp]
    \centering
    \includegraphics[width=0.9\linewidth]{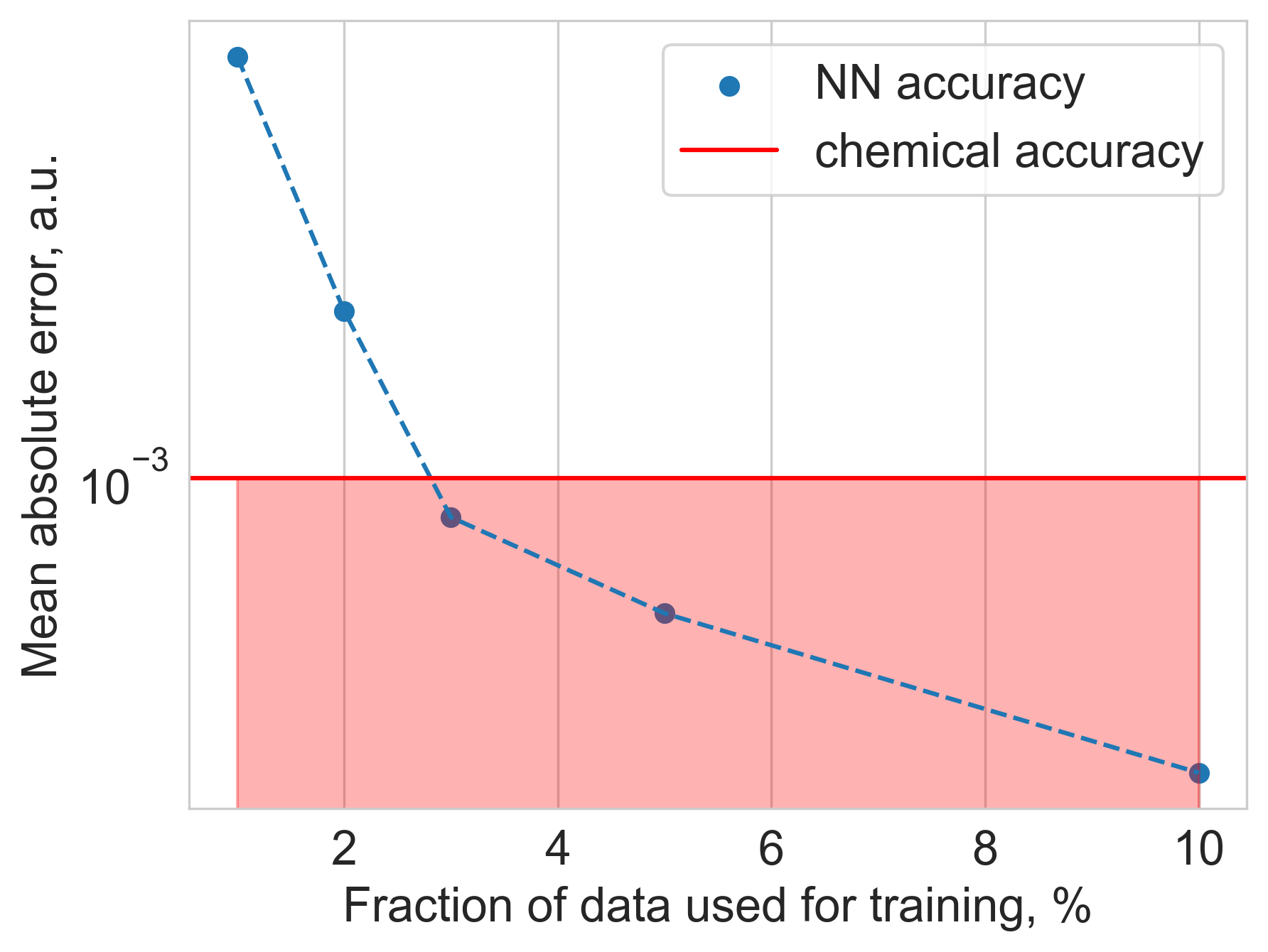}
    \caption{MAE on the test set vs amount of training data}
    \label{fig:chemical_accuracy}
\end{figure}

To see whether the network would be more accurate if there were more training data, we sampled data for \ce{H6} on a finer grid, generating 61,408 structures. Now that we have more data, we can consider a network with more capacity, so we chose a NN that contains 200 neurons for the input layer, followed by 3 layers with 50 neurons, and an output layer with one neuron. As before, the input layer has ReLu activation function and all hidden layers have sigmoid activation function. We also used the same optimizer and scheduler as in our first study of \ce{H6}. Fig. \ref{fig:chemical_accuracy} shows how the mean absolute error (MAE) decreases as the amount of training data increases. As expected, given sufficient training data ($\sim 1500$ structures), the model provides chemical accuracy. The training curve suggests that, given sufficient data, arbitrary accuracy can be achieved. 

There are three questions that need to be addressed about the proposed methods. First, does our descriptor effectively capture the correlation regime, or is it merely another way to model the potential energy surface? The second concern—which is more practical—is how to apply this methodology to larger systems, where generating thousands of training samples is simply infeasible. Finally, is it feasible to use our descriptor in a size independent context, without either explicit or implicit dependence on the number of atoms? We address those questions below.

\begin{figure}
    \centering
    \includegraphics[width=\linewidth]{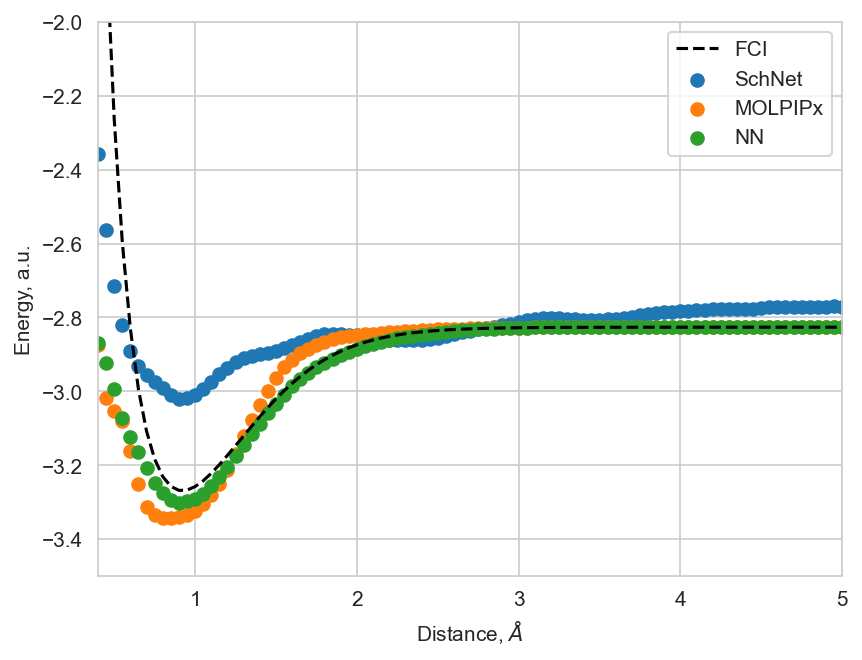}
    \caption{Predicted and FCI energies for the dissociation of the linear \ce{H_6} chain, computed with the STO-6G basis set. The MAE for predicted energies is 0.053 Hartree.}
    \label{fig:h6_transfer}
\end{figure}

To verify that our model learns the energy by modeling different correlation regimes, and is not just a parameterization of the potential energy surface, we tested the descriptor's transferability. We applied our model—trained on families of six-atom hydrogen clusters—to a type of system that was not present in the dataset: stretching a linear 6-hydrogen atom chain. As seen in Fig.\ref{fig:h6_transfer}, the model captures the exact energy for stretched geometries (where there is abundant training data) but fails in the compressed bonding region, where the chain of perfect-pairing structures that dominates correlation is not present in our 2- and 3-dimensional training data. Nonetheless, our descriptor outperforms both SchNet and PIPs trained on the same data (see Table \ref{tab:h6_stretch}). It is clear from Fig.\ref{fig:h6_transfer} that SchNet is significantly overfit and does not capture the correct PES. At the same time PIPs capture the correct qualitative behavior, but are inferior to our descriptor. 

\begin{table}[h]
\centering

\begin{tabular}{|l|S|S|S|}
\hline
 & {SchNet} & {MOLPIPx} & {NN} \\ 
\hline
MAE, a.u & 0.1819058393808002 & 0.13938764554909475 & 0.05299465629663746 \\ \hline
\end{tabular}
\caption{Mean absolute error for the \ce{H6} dissociation energy curve calculated by different methods}
\label{tab:h6_stretch}
\end{table}

The bottleneck in our method is the generation of sufficient amounts of high-quality data. Especially for large systems, generating thousands of FCI-quality results is intractable. For a method like ours to be practical, then, one needs to leverage training data for few-electron systems to predict results for many-electron systems. 


As an example, we predict the energy of \ce{H10} by training a neural network on data from calculations of \ce{H2}, \ce{H4}, \ce{H6}, and \ce{H8}. The key insight is that FCI is size-consistent. For example, for infinitely separated fragments, \ce{H8\bond{...}H2}, the energy is the sum of the energies of the $\ce{H8}$ and $\ce{H2}$ fragments and the Hamiltonian matrix elements, $k_{pqrs}$, are zero whenever the indices refer to matrix elements on two different systems. We generated training data for \ce{H10} from data from 4,560,408 smaller (noninteracting) hydrogen clusters, \ce{H8\bond{...}H2}, \ce{H6\bond{...}H4}, and \ce{H6\bond{...}H2\bond{...}H2}. Then we apply the procedure described in Section II to obtain the input feature vector. 

To predict the energy of \ce{H10}, we start with an initial layer with 500 neurons, followed by two layers with 200 neurons each, and then one layer with 100 neurons. These layers use the ReLU activation function and are followed by an output layer with a single neuron and a linear activation function.

\begin{figure}
    \centering
    \includegraphics[width=\linewidth]{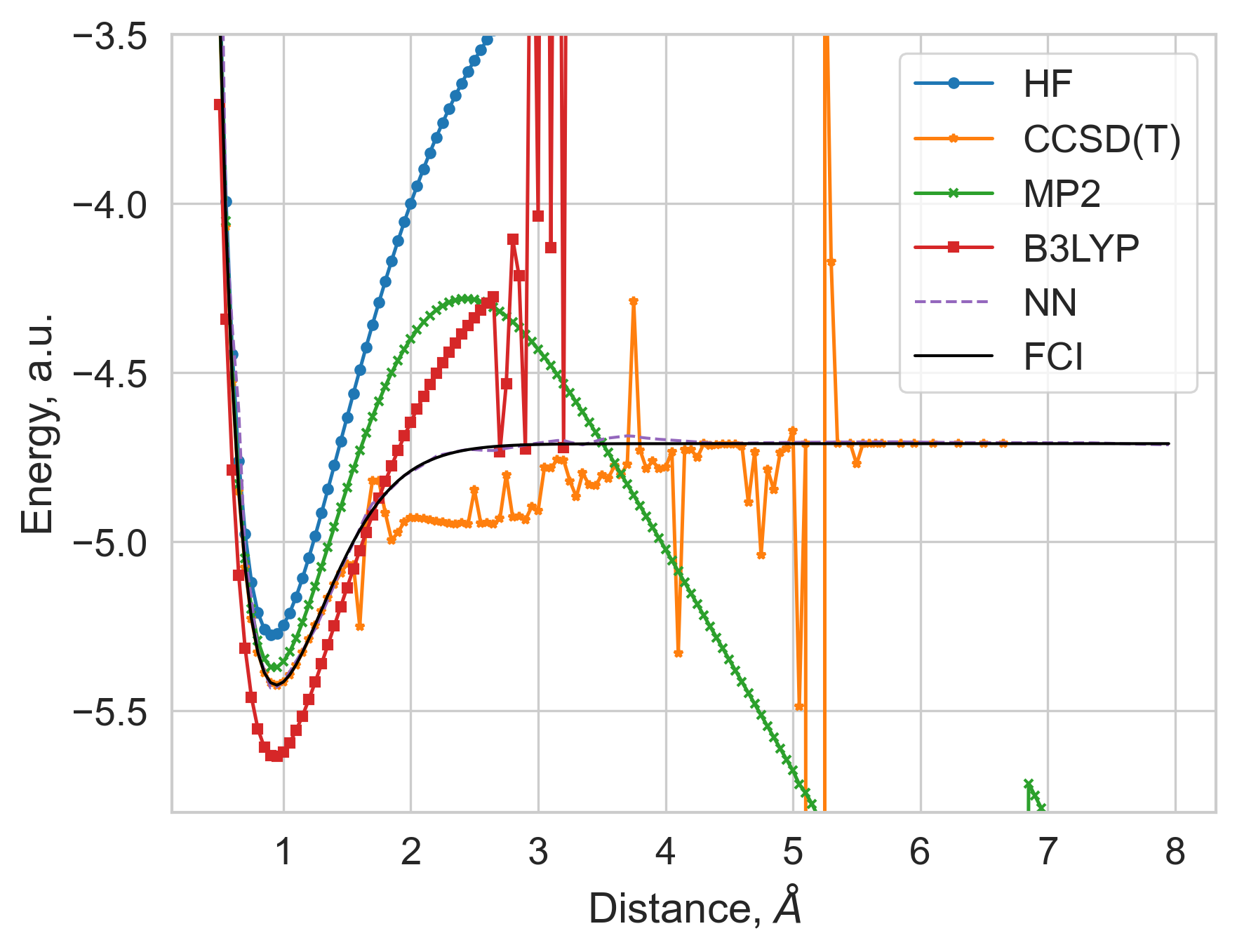}
    \caption{Energy prediction of the dissociation curve for linear chain of ten hydrogen atoms using different methods}
    \label{fig:methods_energy}
\end{figure}

The trained network is then used for fine-tuning. The idea is that the larger network has already learned useful features and transformations of the initial feature vector and can be applied to a real system composed of 10 hydrogen atoms. To achieve this, we unfroze and retrained \textit{only} the first layer of the neural network using just 25 FCI calculations of \ce{H10}. This result gave impressive accuracy for stretching the \ce{H10} chain with a MAE of $\approx 0.01$ a.u.. This is far more accurate than traditional quantum chemistry methods (see Table II). The potential energy surface obtained for different methods is shown at Fig.\ref{fig:methods_energy}: note that the neural network is visually indistinguishable from FCI along the entire potential energy curve. The biggest deviations are for compressed geometries, mainly due to the sparsity of our training data in this regime.

\begin{table}[]
\begin{tabular}{|c|c|c|c|c|c|}
\hline
{\textbf{}} & HF & CCSD(T) & MP2 & B3LYP & NN \\ \hline
{ MAE, a.u} & {1.6267}      & {0.1791}          & { 0.3615}      & {2.4493}         & { 0.0102}      \\ \hline
\end{tabular}
\caption{Mean absolute error for the H10 dissociation energy curve calculated by different methods}
\end{table}

Finally, we demonstrate how the proposed approach can be extended to a size-independent setting. To achieve this, we employ a transformer architecture with single-head self-attention\cite{Vaswani2017, Lee2019}, which naturally accommodates variable input sizes and preserves permutation equivariance. In this framework, the input is represented by a matrix $X \in \mathbf{R}^{N \times d}$, where $N$ denotes the number of elements (e.g., geminals) and $d$ is their embedding dimensionality (just one in our case). Importantly, $N$ can vary across systems, enabling size-independent modeling.

The attention mechanism operates by projecting the input $X$ into three learned representations: queries ( $Q$ ), keys ( $K$ ), and values ( $V$ ). These are obtained via linear transformations parameterized by trainable weight matrices $W_Q, W_K$, and $W_V$, respectively,

$$
\begin{aligned}
K & =X W_K, \\
V & =X W_V, \\
Q & =X W_Q .
\end{aligned}
$$
Here, the weight matrices define the feature subspaces in which similarity comparisons and information aggregation are performed. The self-attention operation is then computed as
$$
\operatorname{Attention}(Q, K, V)=\operatorname{softmax}\left(\frac{Q K^T}{\sqrt{d_k}}\right) V,
$$
where $d_k$ is the dimensionality of the key (and query) vectors. The scaled dot-product $Q K^T / \sqrt{d_k}$ quantifies pairwise interactions between all elements in the input, while the softmax function ensures a normalized weighting over these interactions. The resulting attention weights are used to form a weighted combination of the value vectors, allowing each element to adaptively incorporate information from all other elements. Because this operation depends only on pairwise similarities and not on a fixed system size, the model naturally generalizes to inputs with arbitrary $N$.
It worth noting that we don't use positional encoding, which makes our approach similar to modern set-transformers \cite{Lee2019}.

\begin{figure}
    \centering
    \includegraphics[width=\linewidth]{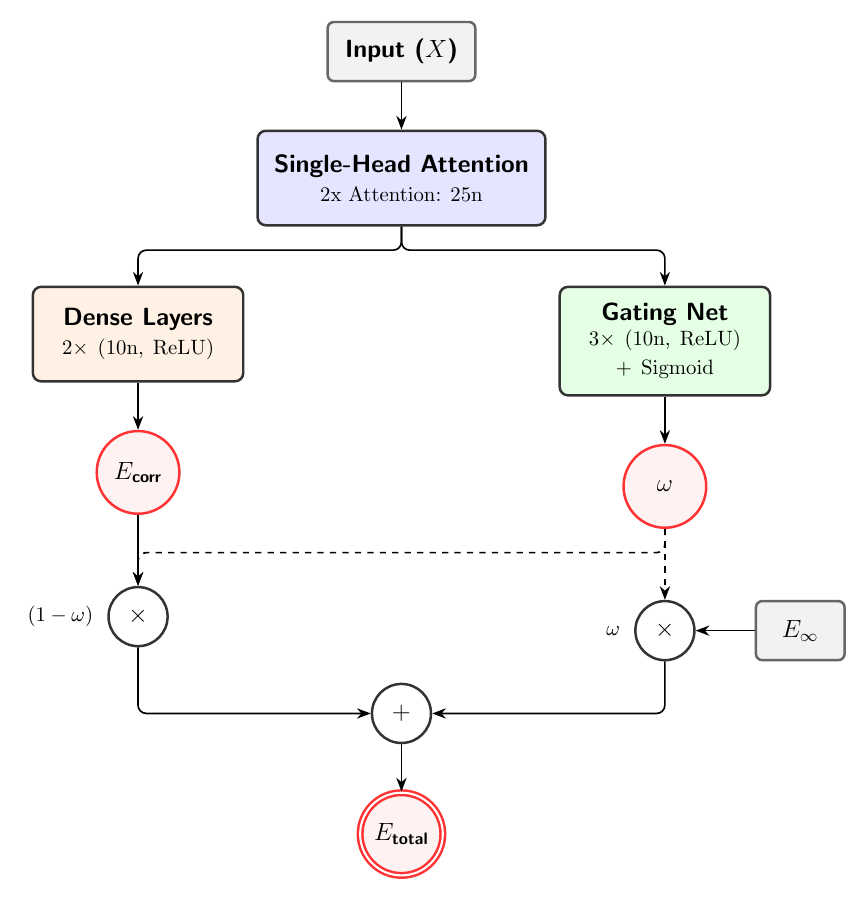}
    \caption{Architecture of the physics-informed Attention mechanism, featuring an asymptotic gate ($\omega$) to enforce physically valid dissociation limits}
    \label{fig:model}
\end{figure}

The proposed architecture contains three primary components: (i) a multi-head attention block for feature extraction, (ii) a dense regression block for the correlation energy $E_{\text {corr }}$, (iii) a physics-informed gating mechanism,  $\omega$, that enforces the correct dissociation limit as an asymptotic boundary condition. The feature extraction stage consists of two sequential multi-head attention layers, enabling global interactions among electron-pairs; the first layer is a single attention head with a dimensionality of $d_k = 25$ neurons each, followed by a second layer with a single head of the same dimensionality. Following the attention block, the network branches into two pathways. The first pathway predicts the correlation energy using two dense layers, each consisting of 10 neurons with ReLU activation. The second pathway acts as a gating mechanism designed to enforce correct behavior at the dissociation limit. This gate $\omega$ is parameterized by a series of three dense layers (10 neurons each, ReLU activation). The final step of the gating mechanism is a sum of values followed by a sigmoid activation function. The gate interpolates between the network's predicted energy and the theoretical dissociation limit, $E_{\infty}$ , which is approximated as half the sum of the eigenvalues of all occupied geminals. Consequently, the total energy is computed as:
$$
E_{\text {total }}=(1-\omega) E_{\text {corr }}+\omega E_{\infty}
$$
where the dissociation energy is computed as $$E_{\infty}=0.5 \sum_i^{N_{\text {occ }}} \varepsilon_i$$
This architecture is shown in Fig. \ref{fig:model}.

\begin{figure}
    \centering
    \includegraphics[width=\linewidth]{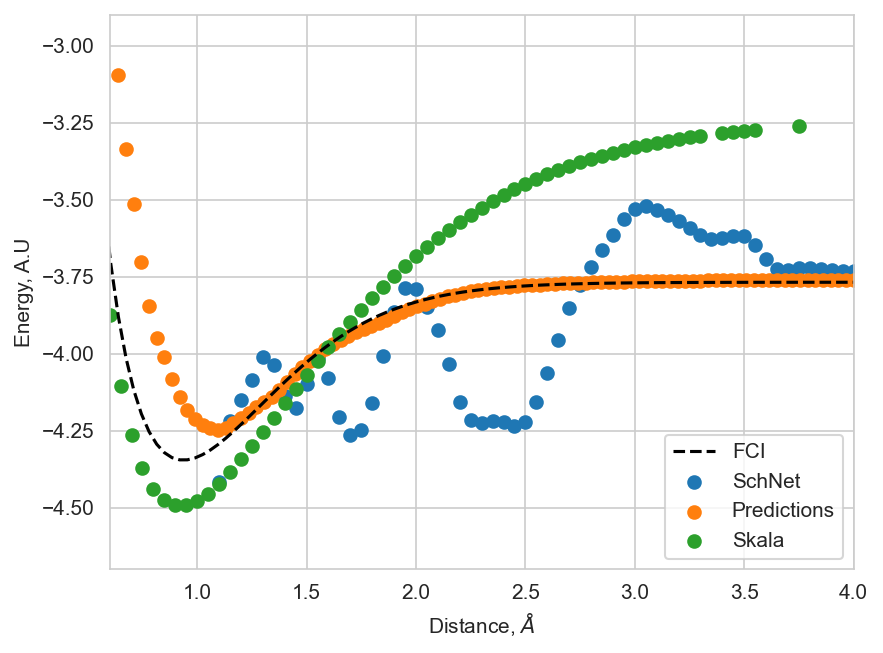}
    \caption{Predicted and FCI energies for the dissociation of the linear \ce{H_8} chain, computed with the STO-6G basis set}
    \label{fig:h8_transformer}
\end{figure}

We trained this model on systems of \ce{H4}, \ce{H6}, and combinations of \ce{H4 \bond{...} H4}, \ce{H6\bond{...} H2}, and \ce{H2 \bond{...} H2 \bond{...} H2 \bond{...} H2} hydrogen clusters. Once trained, we compared performance of the model against SchNet\cite{SchNet} and Skala\cite{Skala}. As before, the SchNet model was trained on the same dataset of \ce{H4} and \ce{H6} hydrogen clusters, while Skala was used as is. As shown in Fig.\ref{fig:h8_transformer} and Table \ref{tab:h8_stretch}, our approaches shows superior stability and generalizability. With an MAE of 0.097, our model achieves a 2.5-fold improvement over SchNet and a 7-fold improvement over Skala and successfully captures both the strong correlation regime and the asymptotic (dissociation) behavior.

Our model reinforces the necessity of aligning neural network architectures with the underlying physics of the system. While SchNet and similar geometric deep learning methods are effective, they rely on local message-passing via learned filters. As aggregation-based graph neural networks, their expressive power is fundamentally constrained by local structural features—a limitation described by the 1-WL framework\cite{xu2018how}. 
Because of this framework, global dependencies must be constructed hierarchically through successive local updates. This bottleneck can limit the model's ability to efficiently capture higher-order structural correlations. This specific inductive bias may account for the performance degradation observed in the out-of-distribution settings of Table \ref{tab:h6_stretch}, particularly in the strongly-correlated regime.

By operating directly on the unitary-invariant geminals' energies with a Transformer architecture, we directly learn the electron correlation regime rather than the local spatial representations inherent in message-passing models. This enables global interactions between electrons, thereby providing a size-independent representation that uses global aggregation of the input features to capture non-local electron correlation.

Although both Skala and SchNet have asymptotic constraints in a form of radial basis functions for convolution filter (SchNet) or explicit cut-off for the learned non-local interactions (Skala), both architectures rely heavily on learned atom-wise representation that can lead to significant overfitting. As shown in Fig.\ref{fig:h8_transformer} this results in unphysical oscillations in SchNet's predicted potential energy surface---or systemic baseline shifts, as seen in Skala's failure to capture the dissociation limit. Conversely, our physics-informed gating mechanism acts as a structural regularizer, anchoring the network to the exact analytical limit ($E_{\infty}$) and ensuring smooth, physically valid, behavior even in the sparse-data regime of stretched geometries.

As expected, the error of our method increases at shorter interatomic distances; this is an artifact of the fact that our model was trained on data built from small hydrogen clusters, with limited data. We could, potentially, impose asymptotic constraints in this regime also.

\begin{table}[h]
\centering
\begin{tabular}{|c|S|S|S|}
\hline
{\textbf{}} & {\textbf{SchNet}} & {\textbf{Skala}} & {\textbf{NN}} \\ \hline
{MAE, a.u} & 0.26004577 & 0.7126474623351507 & 0.09728324545437732 \\ \hline
\end{tabular}
\caption{Mean absolute error for the H8 dissociation energy curve calculated by different methods}
\label{tab:h8_stretch}
\end{table}

\textbf{Summary}:
In order to learn the electronic energy directly from the electronic Hamiltonian, one needs to define input features that capture the symmetries of the quantum-mechanical problem, including rotation and translation invariance of the system and unitary invariance with respect to transformation of the basis set. To do this, we rewrite the electronic Hamiltonian's matrix elements in a geminal basis; the eigenvalues of this matrix are the invariants of our molecular Hamiltonian and have all the information required to determine its ground- and excited-state energies. Using these eigenvalues as the input to a neural network, we can make predictions that are invariant to molecular orientation and the choice of basis. The resulting neural networks are able to achieve chemical accuracy for small hydrogen clusters, an impressive task given the inadequacy of conventional quantum chemistry methods for these systems. The extremely poor accuracy of density-functional and coupled-cluster methods for these systems is unsurprising to quantum chemists, but indicates that traditional approaches where neural networks are trained on data from these computational methods will have large, systematic, potentially incurable errors for some systems. 

To enhance our approach, we introduced a framework for generating training data by combining smaller molecular systems. This approach allows us to pre-train neural networks on a diverse range of few-electron problems and extrapolate to larger chemical systems, where generating training data is very computationally challenging. We used this synthetic data to train a more general, size-independent, transformer model, which, combined with the data from smaller hydrogen clusters outperforms other Neural Networks that utilize geometrical features.  

We believe that our strategy, which combines innovative descriptor design, artificial data generation, transfer learning, and attention mechanisms, is a promising pathway for building deep learning models for large and complex quantum systems. Further refinements---e.g., differentiating the network to obtain reduced density matrices---are clearly possible, and are currently being pursued in our group.

\textbf{Acknowledgments}:\\
The authors acknowledge helpful discussions with Farnaz Heidar-Zadeh (Queen's University) and Rodrigo Alejandro Vargas-Hernandez (McMaster University). Financial support and computational resources were provided by NSERC Discovery and Alliance grants, the Canada Research Chairs, the Digital Research Alliance of Canada, and a Richard Fuller graduate fellowship.

\textbf{References}:
\bibliography{bibliography}

\FloatBarrier
\end{document}


\preprint{AIP/123-QED}
\section{Supplementary Material}

Learning curves for \ce{H4} and \ce{H6} systems along with prediction of dissociation curve of chain of six hydrogen atoms after fine-tuning the model.

\begin{figure}[H] 
    \centering
    \includegraphics[width=0.4\linewidth]{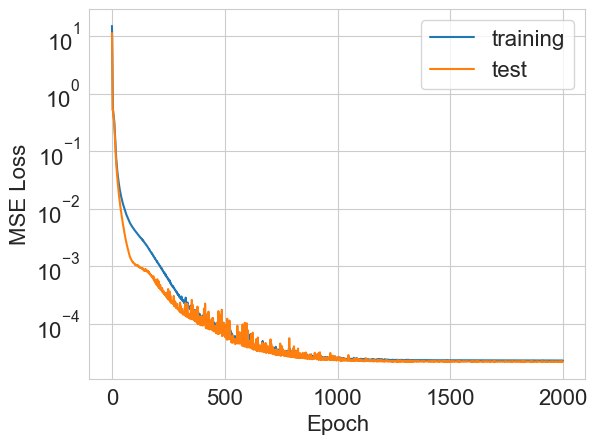}
    \caption{Learning curve of \ce{H4}}
\end{figure}

\begin{figure}[H]
    \centering
    \includegraphics[width=0.4\linewidth]{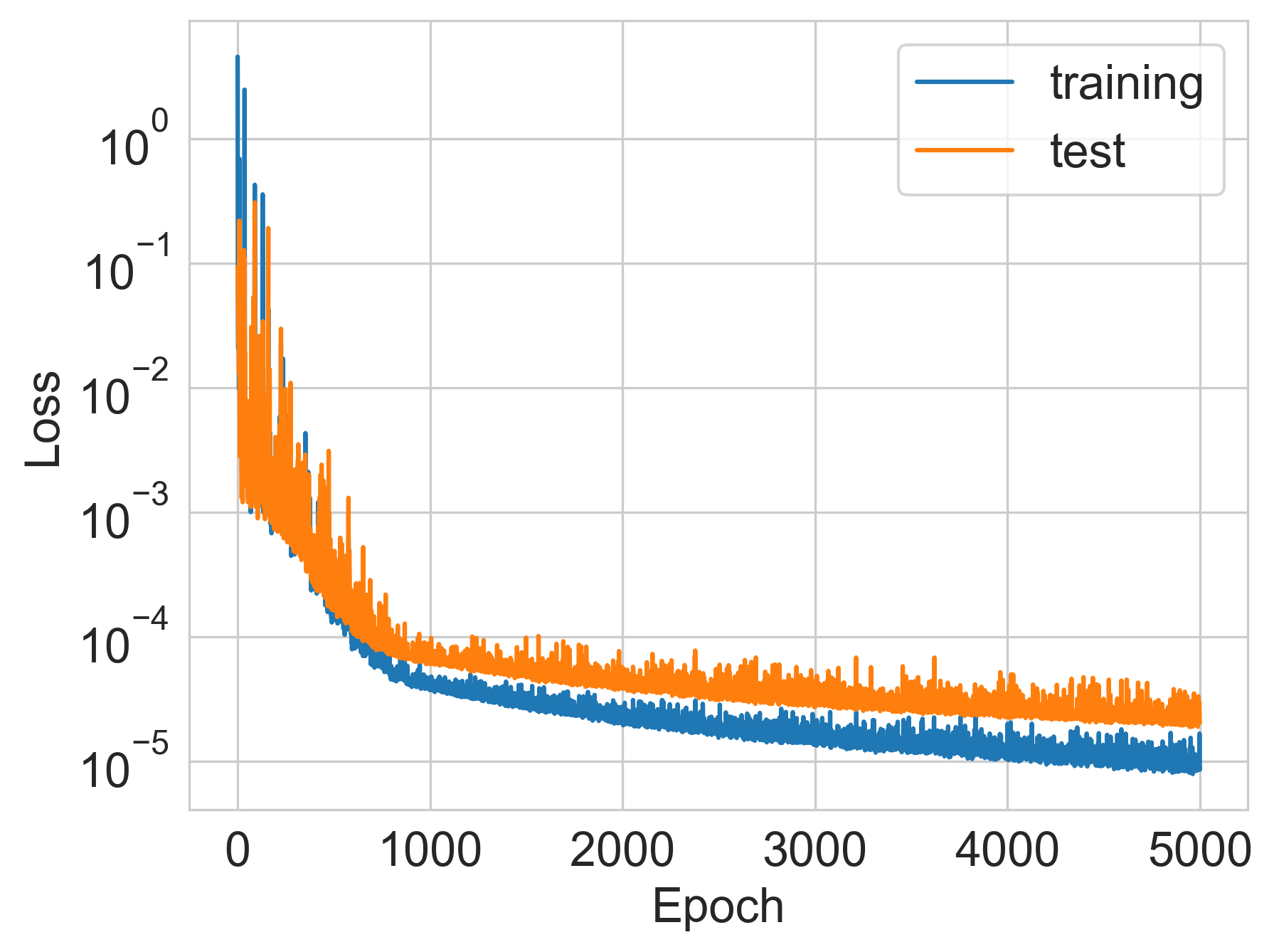}
    \caption{Learning curve of \ce{H6}}
\end{figure}

\begin{figure}[H]
    \centering
    \includegraphics[width=0.4\linewidth]{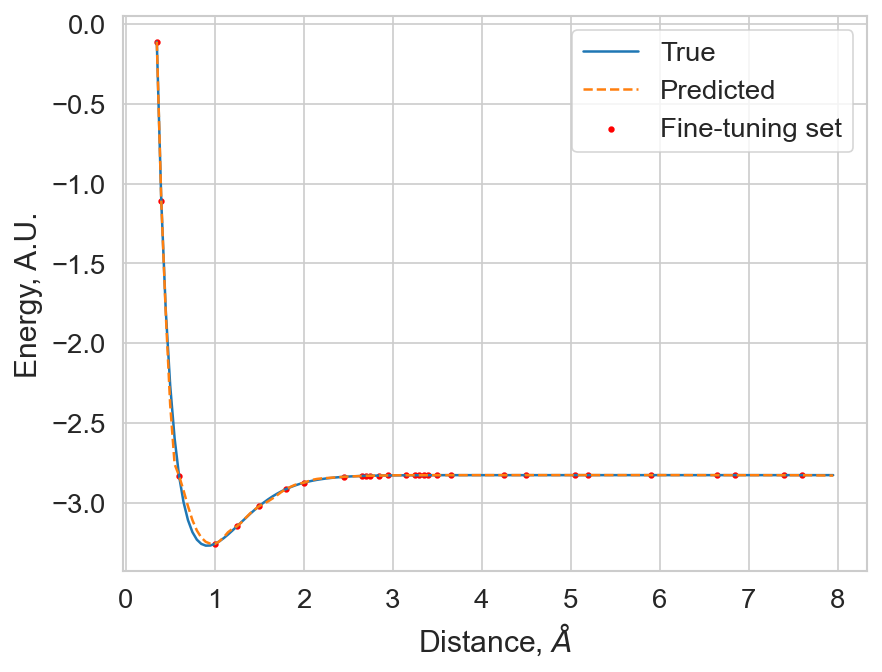}
    \caption{Dissociation curve of \ce{H6}}
\end{figure}

\begin{figure}[htbp]
    \centering
    \includegraphics[width=0.4\linewidth]
    {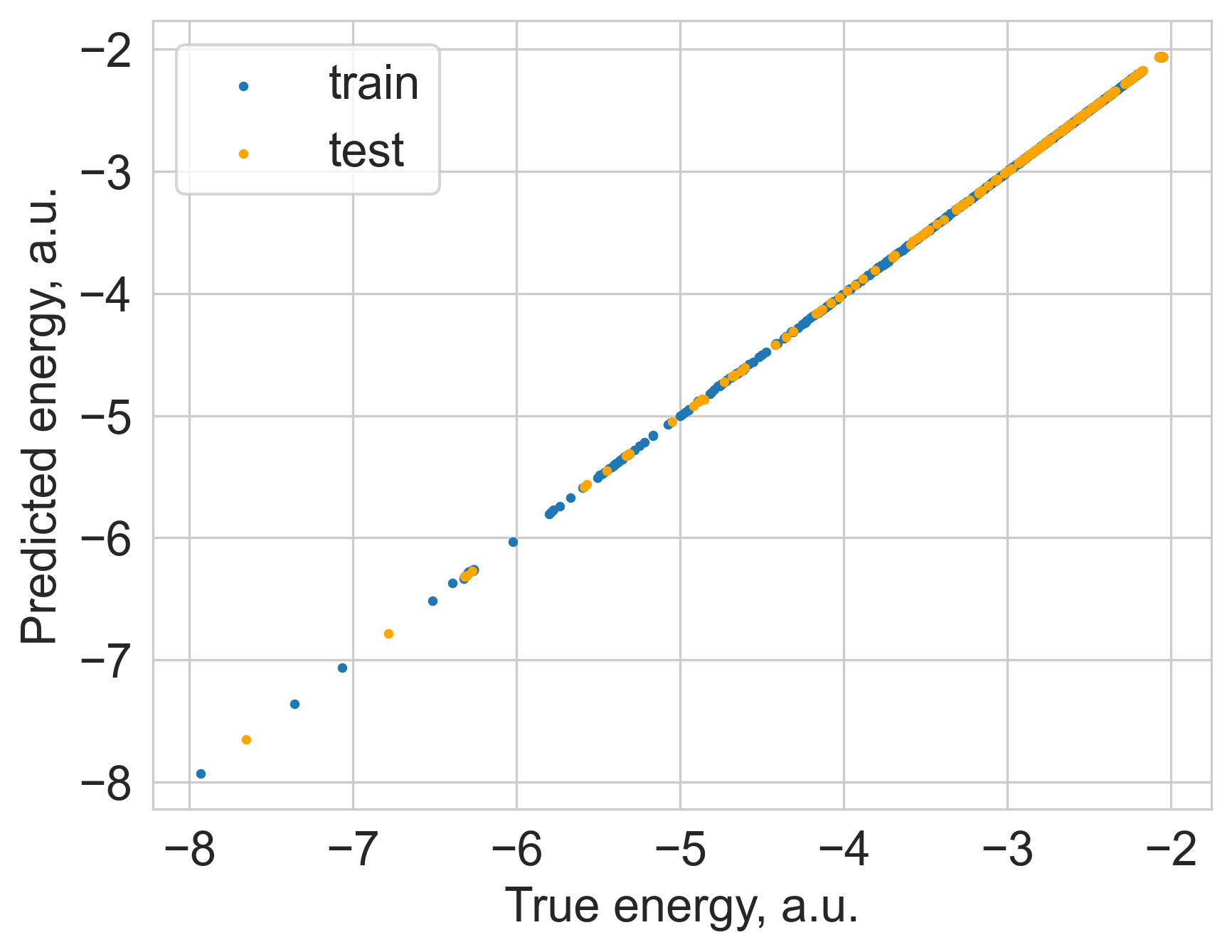}
    \caption{Prediction of training and test data for the H4 systems}
    \label{fig:train_test_h4}
\end{figure}

\begin{figure}[htbp]
    \centering
    \includegraphics[width=0.4\linewidth]{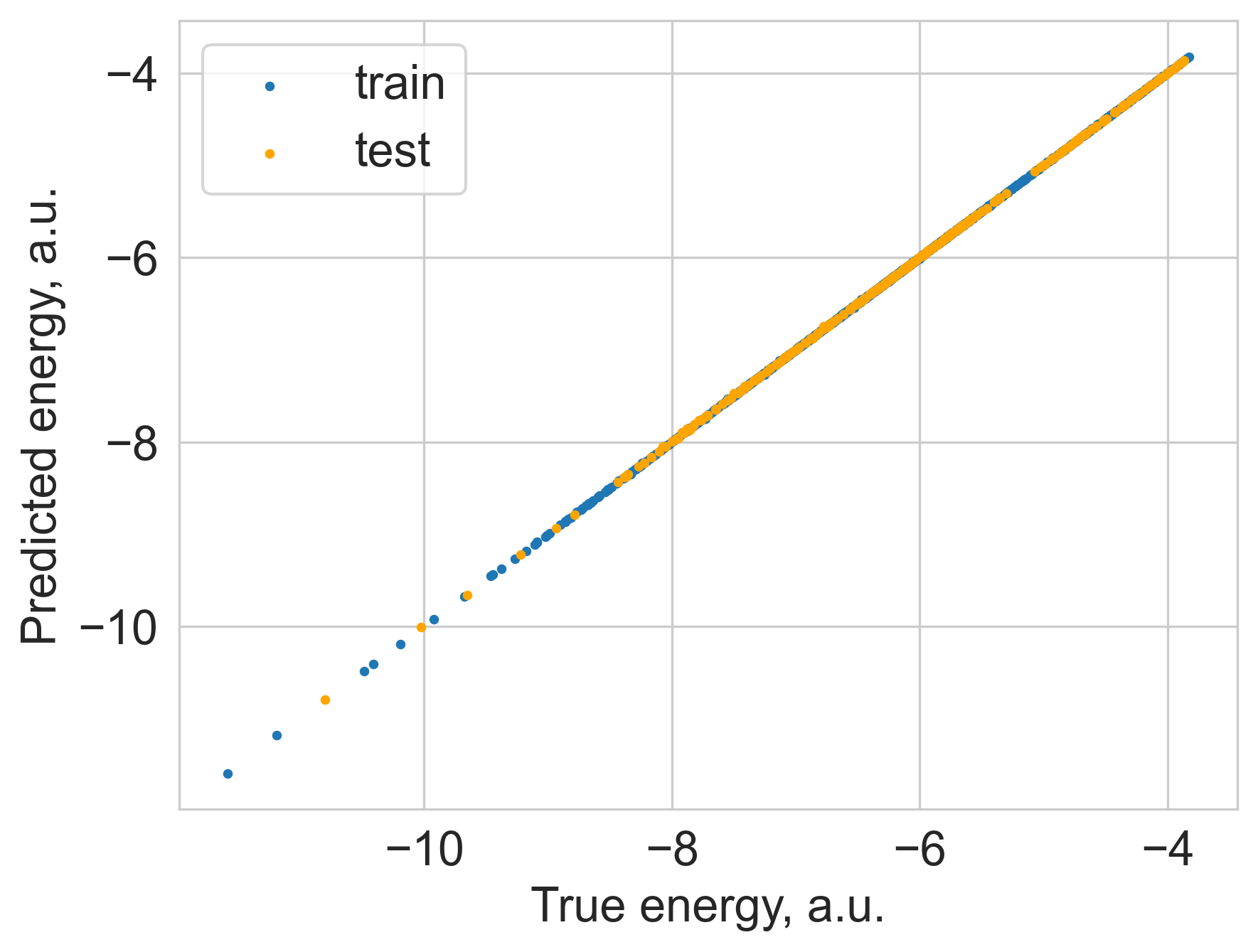}
    \caption{Prediction of training and test data for the H6 systems}
    \label{fig:train_test_h6}
\end{figure}


